# Atomic metasurfaces for manipulation of single photons

Ming Zhou[1], Jingfeng Liu[1,2], Mikhail A. Kats[1,3], Zongfu Yu[1,3]

1. Department of Electrical and Computer Engineering, University of Wisconsin – Madison, 53706, U.S.A.
2. College of Electronic Engineering, South China Agricultural University, Guangzhou 510642, China
3. Department of Material Science and Engineering, University of Wisconsin – Madison, 53706, U.S.A.

**Metasurfaces are an emerging platform for manipulating light on a two-dimensional plane. Existing metasurfaces comprise arrays of optical resonators such as plasmonic antennas or high-index nanorods. In this letter, we describe a new approach to realize metasurfaces based on electronic transitions in two-level systems (TLSs). These metasurfaces can reproduce all of the major results in conventional metasurfaces. In addition, since TLSs can be lossless, tunable, and orders of magnitude smaller than optical resonators, atomic metasurfaces can realize functions that are difficult to achieve with optical resonators. Specifically, we develop quantum scattering theory to describe the interaction between single photons and atomic metasurfaces. Based on the theory, we use a sheet of three-level Λ-type atoms as an example and theoretically demonstrate dynamically tunable single-photon steering. We further show directional beaming of spontaneously emitted photons from atomic metasurfaces.**

Conventional metasurface have seen explosive development in the past few years [1–14]. These optically functionalized surfaces offer completely new ways to control light, and have enabled new phenomena such as anomalous reflection [1], extreme nonlinearity [7], and the spin Hall effect [6], as well as applications such as flat lenses [3], cloaking [11,12] and generation of vortex beams [14].

Figure 1a illustrates the operation principle of conventional metasurfaces [15,16]. A subwavelength optical resonator, for example a nano-rod made from a high-index material [17], imparts a phase onto the incident plane wave. The phase of the scattered light ranges from 0 to π, and is determined by the relative frequency detuning between the incident light and the resonance. Rods with different radii have different resonant frequencies, thereby creating different scattering phases (Fig. 1b).

Optical phase can also be acquired through the scattering process in atomic systems. For example, the resonant scattering from electronic resonances in TLSs can impart a phase on the incident photons. A TLS can capture the energy of a single photon and store it in the excited quantum state for a brief moment, during which a phase is accumulated. The probability amplitude of a single photon scattered by a TLS is calculated and shown in the inset of Fig. 1c. The phase ranges from 0 to π depending on the relative energy detuning $\Delta E/\Gamma = (E - E_{ph})/\Gamma$, where $E$ is the transition energy of the TLS, $\Gamma$ is the energy bandwidth, and $E_{ph}$ is the energy of the incident photon. Figure 1c shows the relation between the scattering phase and $\Delta E/\Gamma$, which closely resembles that of optical resonators (Fig. 1a). An array of TLSs with varying transition energies can create different scattering phases, as shown in Fig. 1d. Similar to the optical resonator array (Fig. 1b), the TLS array can tilt the phase front and steer the directions of reflection and refraction. TLSs can be found in atoms and molecules. Many solid-state implementations are also rapidly emerging, such as quantum dots [18], nitrogen-vacancy centers in diamond [19], and Josephson junctions [20].

Atomic metasurfaces can realize unique features that are extremely challenging for their conventional counterparts. First, atomic metasurfaces are naturally reconfigurable, even in the visible wavelength range. The energy levels of TLSs can be rapidly tuned using laser illumination [21], or external electric [22] or magnetic fields [23]. In contrast, existing reconfigurable metasurfaces rely on incorporating a material with a tunable refractive index into the optical resonators, but achieving a large tuning range has remained a challenge, especially at visible frequencies [24–27]. Second, the scattering elements in atomic metasurfaces can be lossless, with quantum efficiencies in certain TLSs, such as atoms and quantum dots, reaching nearly 100% [28]. Furthermore, their scattering efficiency, measured by the ratio between the optical and physical cross sections, can be extremely large (~10000) [29], 3-4 orders of magnitude larger than that of optical resonators [30]. The extremely compact size of individual elements, which is unattainable in neither dielectric nor metallic optical resonators, is particularly attractive for synthesizing metasurfaces for advanced functionalities such as multi-band operation [31].

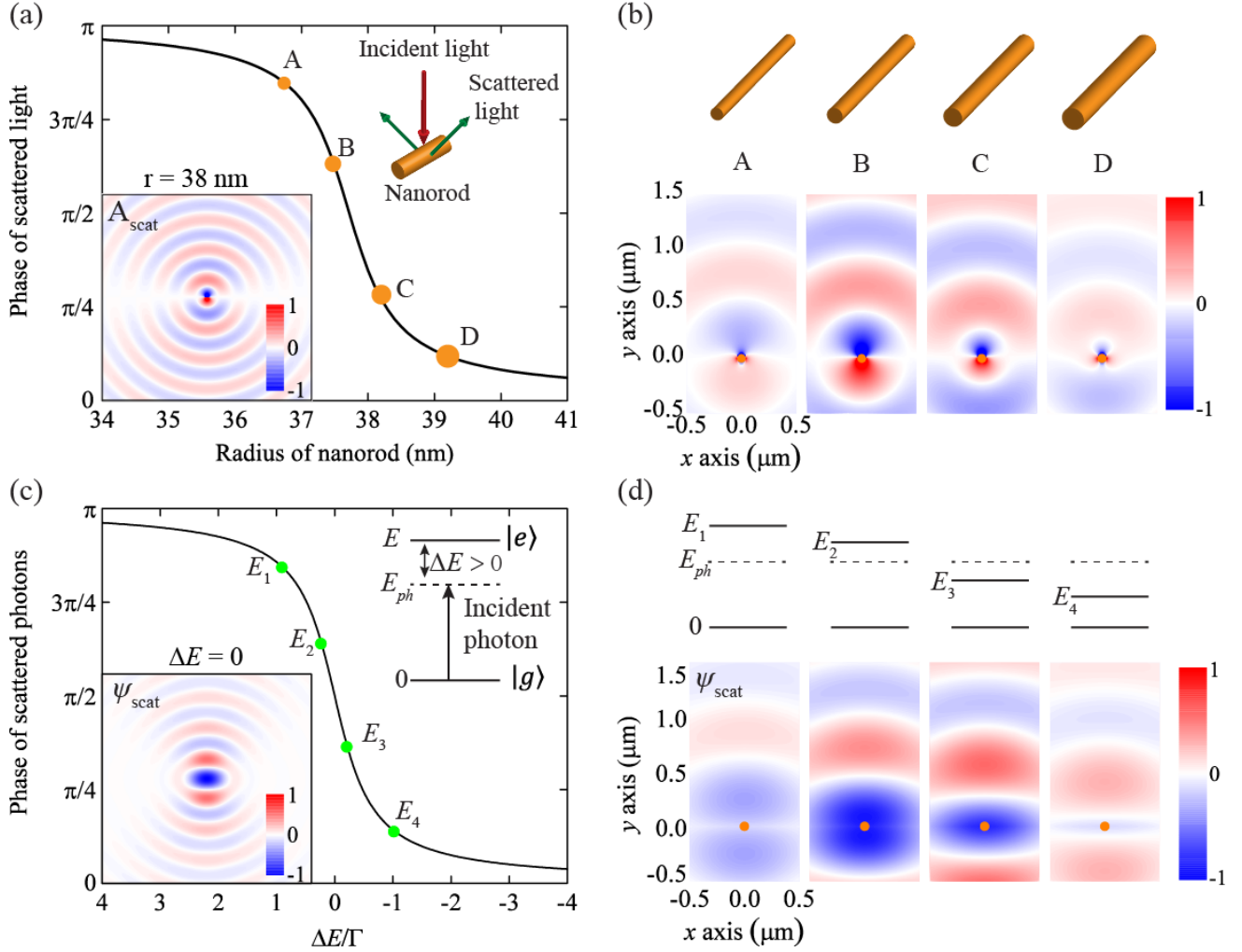

FIG. 1. (a) Scattering phase created by nanorods of different radii. The index of the nanorod is 10 and the incident wavelength is 1 μm. The resonant scattering is caused by the fundamental resonant mode of the nanorod. As the radius of the nanorod increases, the resonant frequency decreases; and the phase of scattered light varies from π to 0. The lower inset shows a typical normalized scattering field calculated from a full-wave simulation. (b) A tilted phase front can be synthesized from light scattered by an array of nanorods with different radii. All fields in (a) and (b) are normalized by the magnitude of incident light. (c) Scattering phase created by TLSs of different transition energies. The energy detuning is normalized by the TLS linewidth Γ. The upper inset shows the energy levels. The lower inset shows the wave function of scattered single photons $\psi_{scat}$ calculated from quantum scattering theory. (d) A tilted phase front can be synthesized from photons scattered by TLSs with different transition frequency. All fields in (c) and (d) are normalized by the magnitude of incident photon.

In this letter, we theoretically demonstrate examples of metasurface that can realized tunable single-photon steering and beaming of spontaneous emission. The properties of an atomic metasurface contains $M$ TLSs are directly studied based on the exact quantum mechanical solution to the Hamiltonian [29]:

$$H = H_0 + H_i \tag{1}$$

The first term is the free Hamiltonian $H_0 = \sum_{m=1}^{M} \hbar\omega_m \sigma_m^\dagger \sigma_m + \sum_k \hbar\omega_k c_k^\dagger c_k$, including all constituent TLSs and the free space photons. Here $\hbar$ is the reduced Planck constant and $\sigma_m^\dagger$ ($\sigma_m$) is the atomic raising (lowering) operator of the $m^{th}$ TLS, which has a transition frequency of $\omega_m$. The creation (annihilation) operator for photons with an angular frequency $\omega_k$ is $c_k^\dagger$ ($c_k$), with $\boldsymbol{k}$ being the momentum of photons. The second term $H_i = i\hbar \sum_{m=1}^{M} \sum_k g_{\boldsymbol{k},m}(c_k^\dagger e^{-i\boldsymbol{k}\cdot\boldsymbol{r}_m}\sigma_m + c_k e^{i\boldsymbol{k}\cdot\boldsymbol{r}_m}\sigma_m^\dagger)$ describes the interaction between the TLSs and photons under the dipole and rotating-wave approximations [29]. The position of the $m^{th}$ TLS is $\boldsymbol{r}_m$, and the coupling strength between the TLS and photons is $g_{\boldsymbol{k},m}$.

To characterize the scattering properties, we recently developed an exact quantum scattering theory based on a non-perturbative Bethe-ansatz method [32]. It uses the Fock states to describe the incident light and directly calculate the free-space eigenstate of the Hamiltonian in Eq. (1). All results shown in this letter are based on exact solutions from the quantum scattering theory and the derivation is available in Sec. 1 Supplemental Material.

We first demonstrate an example of reconfigurable atomic metasurface. A metasurface has to create different scattering phases at different locations. Most conventional metasurfaces achieve the phase variation by using optical resonators with different resonant frequencies, which can be done easily by using resonators of different sizes. In contrast, it would be difficult to find TLSs with their natural energy level continuously varying in a controlled way. To overcome this difficulty, we exploit dressed quantum states in three-level Λ-type system, for example, $^{87}$Rb atoms [33]. The transition energy to a dress state can be easily tuned through laser illumination. A spatially structured illumination can create a spatially varying scattering phase. More importantly, since the resonant transition is controlled by external input, these metasurfaces can be rapidly reconfigured.

The energy level diagram of a three-level Λ-type system is shown in Fig. 2a. When illuminated by a control laser at the same wavelength with the transition between the states |2⟩ and |3⟩, the two states are resonantly coupled, creating two dressed states $|a\rangle = (1/\sqrt{2})(|2\rangle + |3\rangle)$ and $|b\rangle = (1/\sqrt{2})(|2\rangle - |3\rangle)$. The energy levels of the two dressed states are separated by a Rabi frequency $\Omega_c = d_{23}|A_c|/\hbar$, where $d_{23}$ is the transition dipole moment between states |2⟩ and |3⟩, and $|A_c|$ is the magnitude of the electric field of the control laser. A new TLS is formed between |1⟩ and the dressed state |b⟩. Most importantly, its transition energy $E = E_2 - E_1 - \hbar\Omega_c/2$ is dynamically tunable because $\Omega_c$ is linearly proportional to the amplitude of the control laser. Using a control laser with spatially varying illumination, we can create arrays of spatially varying TLSs.

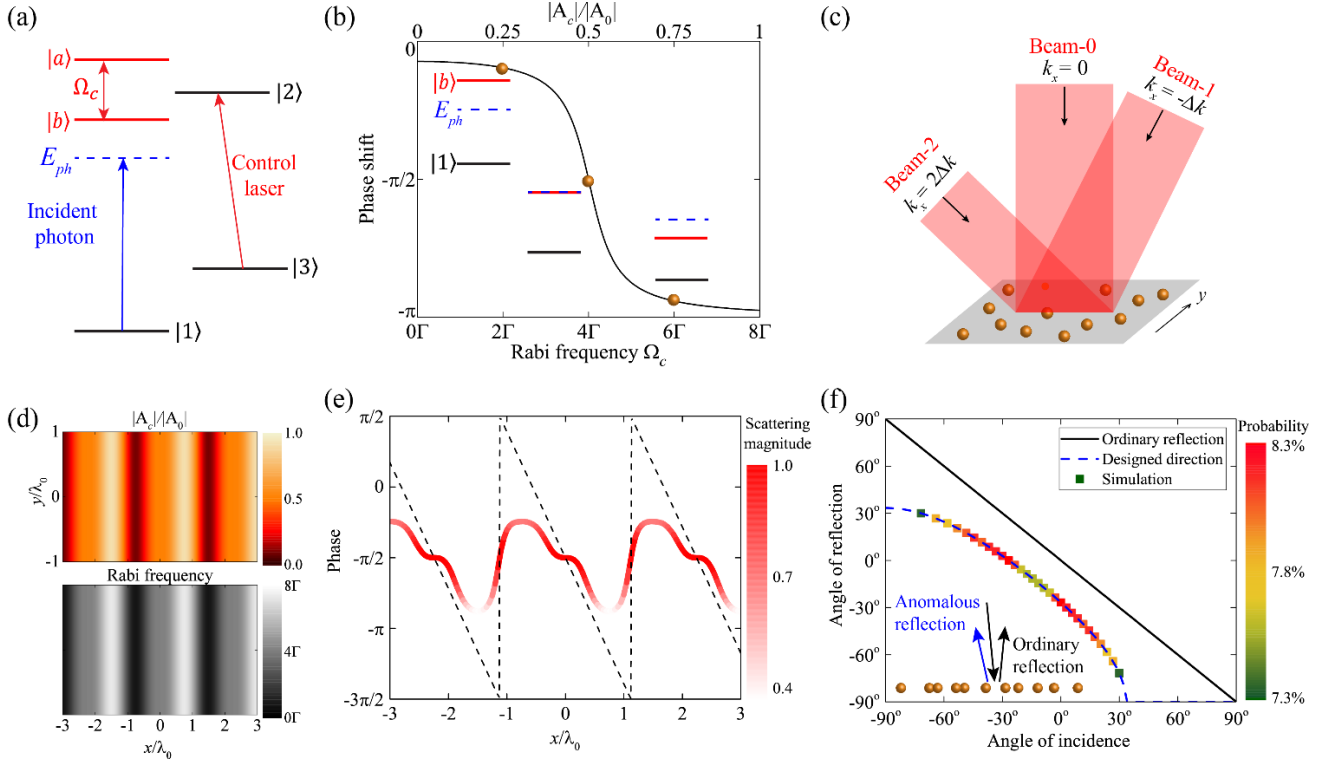

FIG. 2. (a) Energy diagram of three-level atoms. When the states $|2\rangle$ and $|3\rangle$ are coupled by the control laser, two dressed states $|a\rangle$ and $|b\rangle$ are created. The transition between the ground state $|1\rangle$ and the lower dressed state $|b\rangle$ serves as a tunable TLS to be used in an atomic metasurface. (b) The scattering phase for an incident photon with fixed energy as indicated by the blue dashed line. The amplitude of control laser $A_c$ is normalized by $A_0$, which corresponds to a Rabi frequency of $2\Gamma$ and $\Gamma$ is the energy bandwidth between states $|1\rangle$ and $|2\rangle$. As the amplitude of control laser and the Rabi frequency changes, the phase spans a range of approximately $\pi$. (c) The schematic of the metasurface with atoms randomly positioned on the *x-y* plane. The interference of three control beams creates a spatial profile that tunes the Rabi frequencies at different locations. (d) The spatial profile of normalized electric field of the control beams (upper panel) and the Rabi frequency (lower panel). (e) The spatial profile of the scattering phase. The scattering amplitude is encoded by the weight of the red curve. The black dashed line indicates the ideal phase gradient to impart a phase gradient of $\Delta k$ to the incident photons. (f) Incident photons are steered to the designed direction (blue dashed line). Simulation results (square markers) is calculated from the full quantum treatment. The inset shows the schematic of the directions of the incident, ordinary reflection and beam steering.

Figure 2b shows the scattering phase of a single photon scattered by the three-level system. The photon energy $E_{ph}$ is chosen to be close to, but detuned from, the transition energy between $|1\rangle$ and $|2\rangle$ (blue dashed line in Fig. 2a). As an example, we set $E_{ph} = E_2 - E_1 - 2\Gamma$. When the control laser is off, the scattering phase is nearly zero since its frequency is far away from any resonance. As we turn on the control laser and increase its intensity, the Rabi frequency increases and $|2\rangle$ splits into two states. As the energy of the lower dressed state $|b\rangle$ decreases and approaches the resonant energy level $E_1 + E_{ph}$, the scattering phase decreases (Fig. 2b). When $|b\rangle$ sweeps across the resonant level, the phase varies over a range of approximately $\pi$.

Now we can design a reconfigurable metasurface using three-level atoms as building blocks. The atoms are *randomly* placed on a plane. The average density of atoms is 4 atoms per area of $\lambda_0^2$, where $\lambda_0$ is the transition wavelength between states $|1\rangle$ and $|2\rangle$. The metasurface is designed to have a spatial phase gradient $\Delta k = \Delta\phi/\Delta x$ along the $\hat{x}$-axis, which steers the incident single photons away from the ordinary reflection direction (Fig. 2f). To create this phase gradient, we use three beams from the same control laser to generate structured illumination. Figure 2c shows the directions of these beams. Their wave vectors in the $x$ direction are $k_x = 0, -\Delta k, 2\Delta k$, respectively. The resulting interference pattern is plotted in the upper panel in Fig. 2d (more details in Sec.2 Supplemental Material). The Rabi frequency (lower panel) follows the same pattern. Figure 2e shows the corresponding scattering phase at different spatial locations. A $\pi$ phase range is realized. The scattering amplitude is encoded by the weight of the red line in Fig. 2e, and is designed to maximize the scattering strength in the region where it approximates a linear phase gradient.

We solve the full Hamiltonian of the metasurface, which contains over 1600 atoms within an area of $20\lambda_0$ by $20\lambda_0$. The solution takes into account the collective interactions, such as the superradiance effect [34]. It is necessary to perform such full simulation in order to confirm the designs based on the scattering phase of individual atoms. Specifically, we consider a single photon incident from an angle $\theta_i$. The ordinary reflection direction is shown by the black solid line in Fig. 2f with $\theta_r = -\theta_i$. The metasurface is designed to steer the reflection to a different direction [1] with $\theta'_r = \sin^{-1}(\Delta k/k_0 - \sin\theta_i)$. Blue dashed line in Fig. 2f shows the designed direction for the anomalous reflection with a phase gradient of $\Delta k = -0.9\pi/\lambda_0$. The result from full quantum calculation is shown by square markers. They agree very well. Random positioning of the atoms also leads to weak diffusive reflection, which is too small to be shown in the figure. The efficiency of beam steering, which is between 7.3 ~ 8.3% (Fig. 2f), is indicated by the color of the markers. It can be further improved using the same techniques developed for conventional metasurfaces. For example, using a scattering element that provides a $2\pi$ phase range can significantly improve the steering efficiency. Furthermore, a back mirror that blocks the transmission can double the efficiency. We discuss approaches to realize a $2\pi$ phase range in Sec. 3 Supplemental Material. We further provide examples of metasurfaces with steering efficiency of 98.1% and multiband operation in Secs. 4-5 Supplemental Material.

Next, we demonstrate the dynamic steering of single photons by tuning the directions of the control beams as shown in Fig. 3a. The phase gradient $\Delta k$ is tuned from $-0.9\pi/\lambda_0$ to $0.9\pi/\lambda_0$, resulting in different steering directions as shown by the blue arrows in Fig. 3a. To realize this tuning range, beam-0 is always incident from the normal direction. The incident angle of the 1st beam $\theta_1$ changes from -27° to 27° counter clockwise, while that of the 2nd beam $\theta_2$ changes from 64° to -64° clockwise. Figure 3b shows the results when the reflected single photons are dynamically steered from $-20°$ to $33°$ as the control beams vary. The color of the markers in Fig. 3b indicate the probability of the scattered single photon in the anomalous direction. This steering efficiency remains about the same for different steering angles.

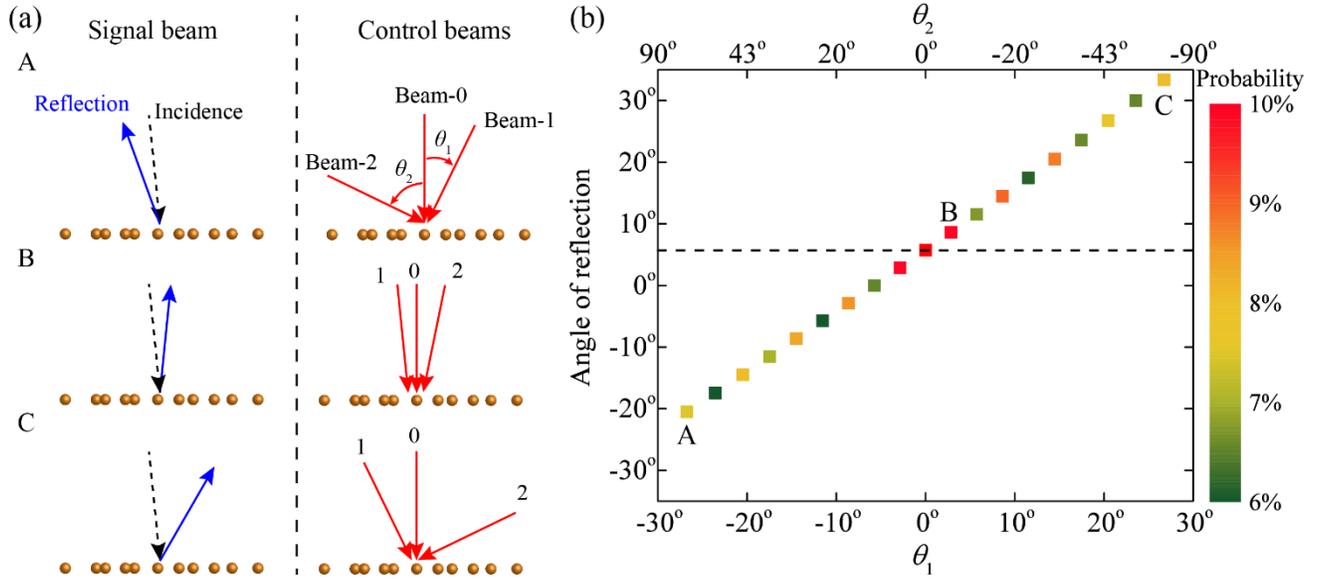

FIG. 3. (a) The signal beam (blue arrows in the left panels) is steered to different directions depending on the s indicated by the dashed line. Square markers represent the angles and the efficiency of the steered beam. Results are calculated from the quantum scattering theory.

Controlling the emission of single photons is also of particular importance in quantum information. Existing methods mostly rely on modifying the local optical density of states [35–37], which involves complex nanostructures. Here we propose to use atomic metasurfaces to realize directional spontaneous emission to a tunable direction. It exploits TLSs' collective interactions to enhance the spatial coherence of the spontaneous emission. Reminiscent of the super- and sub- radiant effects [34], the collective interaction in atomic metasurfaces, however, can be engineered with complex phase configuration to allow unprecedented control of spontaneous emission.

To investigate the spontaneous emission in a metasurface, we solve the time-dependent Schrodinger equation to obtain the temporal evolution of the wavefunction when the initial condition of the metasurface is set to have one TLS in the excited state and no incident photons. To make the calculation feasible on our limited computing facility, we study the case of two-dimensional space, which could be realized by guided waves in planar structure [38]. The working principle is the same in three dimension.

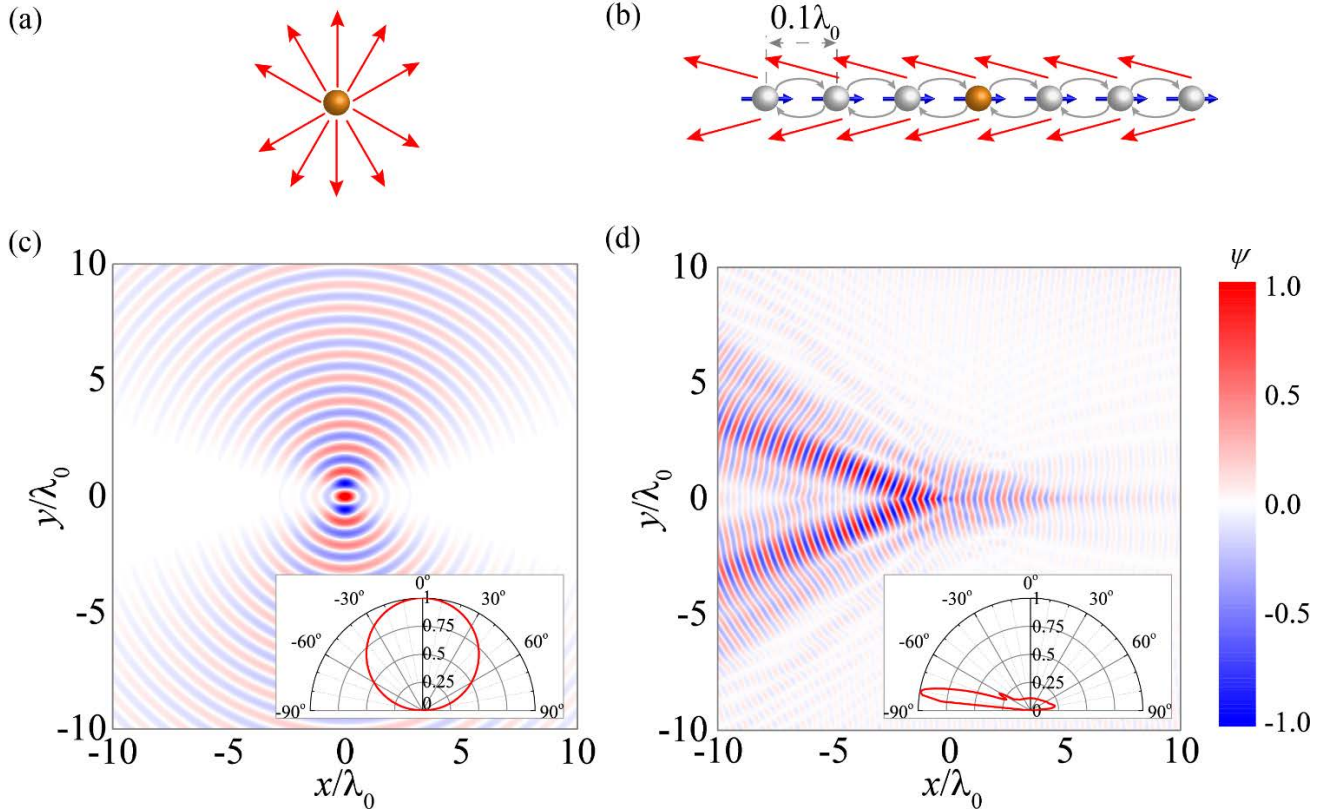

FIG. 4. (a) A single excited TLS in free space. (b) A single excited TLS (golden sphere) in a metasurface. All the other TLSs (gray sphere) are initially in ground state. The total length of the metasurface is $40\lambda_0$, with an inter-TLS spacing of $0.1\lambda_0$, where $\lambda_0$ is the resonant wavelength of the excited TLS. All TLSs have the same lifetime $\tau$. (c) and (d) The snapshot of the real part of wave function at time $t = 3\tau$ for a single excited TLS in free space (c) and within the metasurface (d). The insets show the angular distributions of total spontaneous emission integrated over time.

A single excited TLS in the free space emits with a dipole radiation profile (Fig. 4a). The same TLS in a metasurface (Fig. 4b), however, emits very differently, as shown in Fig. 4b. Only one TLS, marked by golden color, is in the excited state and all other TLSs of the metasurface are in the ground state (Fig. 4b). TLSs have different transition energies, configured to scatter photons with different phases (more details in Sec. 6 Supplemental Material). The snapshot of the wave functions of the embittered photon at $t = 3\tau$ are both plotted the free space (Fig. 4c) and the metasurface (Fig. 4d). A strong directional emission profile is achieved. All TLSs in the metasurface are partially excited through far-field radiative interactions and contribute to the interference that help to steer the emitted single photon. The total flux of the probability integrated over time is shown as insets in Fig. 4c and d. In contrast to the dipolar profile from a single TLS in free space (Fig. 4c inset), the spontaneous emission of a metasurface is highly directional. More important, this direction can be controlled by the external inputs by tuning the phase gradient of the metasurface.

In conclusion, we show that electronic transition can replace the optical resonator as the building block for metasurfaces. As an example, we describe the designs based on three-level atoms. These atomic metasurfaces are quite feasible in experiments, as demonstrated in a previous experimental work that uses a cluster of atoms and structured illumination as diffraction gratings [39]. We also point out that the potential of atomic metasurfaces goes far beyond the control and the generation of single photons. When illuminated with multi-photon states, atomic metasurfaces can exhibit new dynamics and phenomena that are completely absent in conventional metasurfaces. The most significant is the strong photon-photon interactions induced by the Fermionic excitation of TLS. Such interactions could make the atomic metasurface an effective antenna to receive, transmit, and create entangled photons, leading to possible applications in quantum information networks. Because of the strong photon-photon and the tunability of such interaction through external inputs, atomic metasurfaces could also become a versatile platform to study the photonic analog of many-body physics.

**Acknowledgement**: The authors acknowledge the financial support by the Office of Naval Research under Grant No. N00014-14-1-0300.

# Reference

[1] N. Yu, P. Genevet, M. A. Kats, F. Aieta, J.-P. Tetienne, F. Capasso, and Z. Gaburro, Science **334**, 333 (2011).
[2] Y. Zhao and A. Alù, Phys. Rev. B **84**, 205428 (2011).
[3] F. Aieta, P. Genevet, M. A. Kats, N. Yu, R. Blanchard, Z. Gaburro, and F. Capasso, Nano Lett. **12**, 4932 (2012).
[4] S. Sun, K.-Y. Yang, C.-M. Wang, T.-K. Juan, W. T. Chen, C. Y. Liao, Q. He, S. Xiao, W.-T. Kung, G.-Y. Guo, L. Zhou, and D. P. Tsai, Nano Lett. **12**, 6223 (2012).
[5] X. Ni, N. K. Emani, A. V. Kildishev, A. Boltasseva, and V. M. Shalaev, Science **335**, 427 (2012).
[6] X. Yin, Z. Ye, J. Rho, Y. Wang, and X. Zhang, Science **339**, 1405 (2013).
[7] J. Lee, M. Tymchenko, C. Argyropoulos, P.-Y. Chen, F. Lu, F. Demmerle, G. Boehm, M.-C. Amann, A. Alù, and M. A. Belkin, Nature **511**, 65 (2014).
[8] C. Wu, N. Arju, G. Kelp, J. A. Fan, J. Dominguez, E. Gonzales, E. Tutuc, I. Brener, and G. Shvets, Nat. Commun. **5**, 3892 (2014).
[9] D. Lin, P. Fan, E. Hasman, and M. L. Brongersma, Science **345**, 298 (2014).
[10] G. Zheng, H. Mühlenbernd, M. Kenney, G. Li, T. Zentgraf, and S. Zhang, Nat. Nanotechnol. **10**, 308 (2015).
[11] X. Ni, Z. J. Wong, M. Mrejen, Y. Wang, and X. Zhang, Science **349**, 1310 (2015).
[12] L. Y. Hsu, T. Lepetit, and B. Kante, Prog. Electromagn. Res. **152**, 33 (2015).
[13] M. Decker, I. Staude, M. Falkner, J. Dominguez, D. N. Neshev, I. Brener, T. Pertsch, and Y. S. Kivshar, Adv. Opt. Mater. **3**, 813 (2015).
[14] E. Maguid, I. Yulevich, D. Veksler, V. Kleiner, M. L. Brongersma, and E. Hasman, Science aaf3417 (2016).
[15] F. Falcone, T. Lopetegi, M. A. G. Laso, J. D. Baena, J. Bonache, M. Beruete, R. Marqués, F. Martín, and M. Sorolla, Phys. Rev. Lett. **93**, 197401 (2004).
[16] C. L. Holloway, E. F. Kuester, J. A. Gordon, J. O'Hara, J. Booth, and D. R. Smith, IEEE Antennas Propag. Mag. **54**, 10 (2012).
[17] L. Cao, J. S. White, J.-S. Park, J. A. Schuller, B. M. Clemens, and M. L. Brongersma, Nat. Mater. **8**, 643 (2009).
[18] A. Zrenner, E. Beham, S. Stufler, F. Findeis, M. Bichler, and G. Abstreiter, Nature **418**, 612 (2002).
[19] I. Aharonovich, S. Castelletto, D. A. Simpson, A. Stacey, J. McCallum, A. D. Greentree, and S. Prawer, Nano Lett. **9**, 3191 (2009).
[20] O. Astafiev, A. M. Zagoskin, A. A. Abdumalikov, Y. A. Pashkin, T. Yamamoto, K. Inomata, Y. Nakamura, and J. S. Tsai, Science **327**, 840 (2010).
[21] K.-J. Boller, A. Imamolu, and S. E. Harris, Phys. Rev. Lett. **66**, 2593 (1991).
[22] P. Tamarat, T. Gaebel, J. R. Rabeau, M. Khan, A. D. Greentree, H. Wilson, L. C. L. Hollenberg, S. Prawer, P. Hemmer, F. Jelezko, and J. Wrachtrup, Phys. Rev. Lett. **97**, 83002 (2006).
[23] M. Trepanier, D. Zhang, O. Mukhanov, and S. M. Anlage, Phys. Rev. X **3**, 41029 (2013).
[24] Y. Yao, R. Shankar, M. A. Kats, Y. Song, J. Kong, M. Loncar, and F. Capasso, Nano Lett. **14**, 6526 (2014).
[25] Y. Fan, N.-H. Shen, T. Koschny, and C. M. Soukoulis, ACS Photonics **2**, 151 (2015).
[26] Y.-W. Huang, H. W. H. Lee, R. Sokhoyan, R. A. Pala, K. Thyagarajan, S. Han, D. P. Tsai, and H. A. Atwater, Nano Lett. (2016).
[27] Q. Wang, E. T. F. Rogers, B. Gholipour, C.-M. Wang, G. Yuan, J. Teng, and N. I. Zheludev, Nat. Photonics **10**, 60 (2016).
[28] P. Michler, A. Kiraz, C. Becher, W. V. Schoenfeld, P. M. Petroff, L. Zhang, E. Hu, and A. Imamoglu, Science **290**, 2282 (2000).
[29] Z. Ficek and S. Swain, *Quantum Interference and Coherence: Theory and Experiments* (Springer Science & Business Media, 2005).

[30] C. F. Bohren and D. R. Huffman, *Absorption and Scattering of Light by Small Particles* (1983).
[31] F. Aieta, M. A. Kats, P. Genevet, and F. Capasso, Science **347**, 1342 (2015).
[32] J. Liu, M. Zhou, and Z. Yu, Opt Lett **41**, 4166 (2016).
[33] Y. Li and M. Xiao, Phys. Rev. A **51**, R2703 (1995).
[34] R. H. Dicke, Phys. Rev. **93**, 99 (1954).
[35] S. Noda, M. Fujita, and T. Asano, Nat. Photonics **1**, 449 (2007).
[36] C. Belacel, B. Habert, F. Bigourdan, F. Marquier, J.-P. Hugonin, S. Michaelis de Vasconcellos, X. Lafosse, L. Coolen, C. Schwob, C. Javaux, B. Dubertret, J.-J. Greffet, P. Senellart, and A. Maitre, Nano Lett. **13**, 1516 (2013).
[37] W. D. Newman, C. L. Cortes, and Z. Jacob, JOSA B **30**, 766 (2013).
[38] S. Fan and J. D. Joannopoulos, Phys. Rev. B **65**, 235112 (2002).
[39] G. C. Cardoso and J. W. R. Tabosa, Phys. Rev. A **65**, 33803 (2002).

# Supplemental Material

## Atomic metasurfaces for manipulation of single photons

Ming Zhou[1], Jingfeng Liu[1,2], Mikhail A. Kats[1,3], Zongfu Yu[1,3]

1. Department of Electrical and Computer Engineering, University of Wisconsin – Madison, 53706, U.S.A.
2. College of Electronic Engineering, South China Agricultural University, Guangzhou 510642, China
3. Department of Material Science and Engineering, University of Wisconsin – Madison, 53706, U.S.A.

## 1. Quantum scattering theory

As we described in the main text, the general Hamiltonian governing the interaction between photons and *M* TLSs is given by [1]

$$H = \sum_{m=1}^{M} \hbar\omega_m \sigma_m^\dagger \sigma_m + \sum_k \hbar\omega_k c_k^\dagger c_k + i\hbar \sum_{m=1}^{M} \sum_k g_{k,m}(c_k^\dagger e^{-i\boldsymbol{k}\cdot\boldsymbol{r}_m}\sigma_m - c_k e^{i\boldsymbol{k}\cdot\boldsymbol{r}_m}\sigma_m^\dagger) \qquad (S1)$$

Here $\hbar$ is the reduced Planck constant and $\sigma_m^\dagger$ ($\sigma_m$) is the raising (lowering) atomic operator of the $m^{\text{th}}$ TLS. The transition frequency and position of the $m^{\text{th}}$ TLS are $\omega_m$ and $\boldsymbol{r_m}$, respectively. The creation (annihilation) operator for photons with angular frequency $\omega_k$ is $c_k^\dagger(c_k)$, with $\boldsymbol{k}$ being the momentum. The coupling strength between the $m^{\text{th}}$ TLS and the photon with momentum $\boldsymbol{k}$ is given by $g_{k,m}$.

Our strategy for solving the above Hamiltonian is to convert the 3-dimension (3D) problem to a set of 1-dimension (1D) problems, which has been successfully solved in waveguide quantum electrodynamics (QED) [2]. The spatial wavefunctions of single photons then can be explicitly calculated by solving the Schrodinger equation. In this section, we will clearly derive the real-space Hamiltonian and eigenstate step by step. The steady-state solution of the Schrodinger equation will be discussed also.

The key idea of our quantum scattering theory is to convert the summation over the 3D $\boldsymbol{k}$-space to a summation over a set of 1D channels, or "waveguides". Each channel carries a plane wave with a distinct momentum $\boldsymbol{k}$ in a particular direction. They can be illustrated in the $\boldsymbol{k}$-space as shown in Fig. S1a. Here, we use box quantization for the $\boldsymbol{k}$-space by setting up a periodic boundary condition in both *x* and *y* directions. At the end of derivation, we will take the periodicity $L$ to ∞ to effectively remove the impact of this artificial periodic boundary condition. Because of the periodicity, incident photons can only be scattered to a set of discretized channels. These channels are defined by the wave's in-plane wave vectors $\boldsymbol{k_{xy}}$, which are discretized by $\Delta k = 2\pi/L$ (Fig. S1b). For resonance frequency $\omega_0$, the channels are located within a circle of radius $k_0 = \omega_0/c$, where *c* is the speed of light. The total number of channels is $N = \pi \lfloor L/\lambda_0 \rfloor^2$, where $\lambda_0 = 2\pi/k_0$ is the resonance wavelength. The floor operator $\lfloor A \rfloor$ gives the largest integer smaller than $A$.

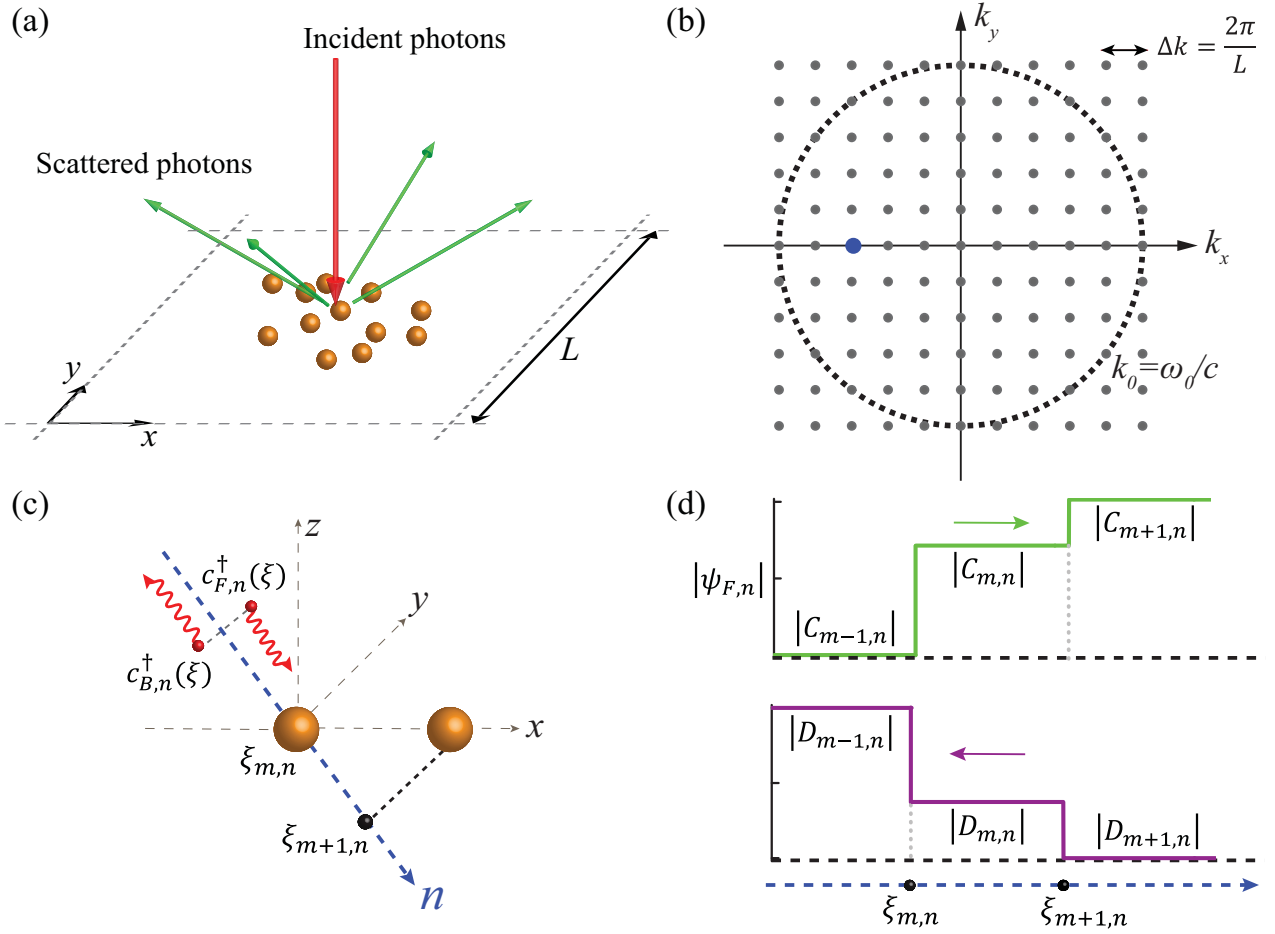

FIG. S1. (a) Schematic of an atomic metasurface containing $M$ TLSs. Periodic boundary conditions with periodicity $L$ are applied in both $x$ and $y$ directions. Because of the periodicity, incident photons can only be scattered to a set of discretized directions. (b) The $\mathbf{k}_{xy}$ space is discretized by $\Delta k = 2\pi/L$, and represents a set of plane-wave channels. The allowed channels are located within the circle with radius $k_0$. (c) Real space bosonic creation operators. The real-space operators $c^\dagger_{F,n}(\xi)$ and $c^\dagger_{B,n}(\xi)$ create a forward and backward propagating single photon at position $\xi$ in the $n^{th}$ channel, respectively. $\xi_{m,n}$ is the position of $m^{th}$ TLS projected along the $n^{th}$ channel. (d) Spatial single-photon wavefunctions $\psi_{F,n}(\xi)$ and $\psi_{B,n}(\xi)$ in the $n^{th}$ channel. The coefficients $C_{m,n}$ and $D_{m,n}$ are the amplitudes of forward and backward propagating photons between $m^{th}$ and $m+1^{th}$ TLSs in $n^{th}$ channel, respectively. These coefficients describe the interaction between TLSs.

Thereby we convert the summation over the 3D $\mathbf{k}$-space to

$$\sum_{\mathbf{k}} \hbar\omega_{\mathbf{k}} c^\dagger_{\mathbf{k}} c_{\mathbf{k}} = \sum_{n=1}^{N} \sum_{k} \hbar\omega_k c^\dagger_{k,n} c_{k,n} \tag{S2}$$

$$\sum_k g_{k,m}(c_k^\dagger e^{-i\mathbf{k}\cdot\mathbf{r}_m}\sigma_m - c_k e^{i\mathbf{k}\cdot\mathbf{r}_m}\sigma_m^\dagger) = \sum_{n=1}^N \sum_k g_{k,m}(c_k^\dagger e^{-ik\xi_{m,n}}\sigma_m - c_k e^{ik\xi_{m,n}}\sigma_m^\dagger) \quad (S3)$$

where $\xi_{m,n}$ is the position of the $m^{th}$ TLS projected along the $n^{th}$ channel. The coupling coefficient $g_{k,m}$ is given by $g_{k,m} = \omega_m\sqrt{1/2\hbar\varepsilon_0\omega_k V}\,\mathbf{d}_m \cdot \mathbf{e}_n$, where $\varepsilon_0$ is the vacuum permittivity, $\mathbf{d}_m$ is the dipole moment of the $m^{th}$ TLS and $\mathbf{e}_n$ is the polarization of photons in the $n^{th}$ channel. The normalization volume $V$ is given by $V = L^3$, where $L$ is the periodicity described in Fig. S1a. Note the wavenumber $k$ is a scalar. It's also useful to differentiate the forward ($F$) and backward ($B$) propagating plane-waves in each channel. Therefore, we define two scalar wave numbers $k_F$ and $k_B$ and get

$$\sum_k \hbar\omega_k c_k^\dagger c_k = \sum_{n=1}^N \left(\sum_{k_F} \hbar\omega_{k_F} c_{k_F,n}^\dagger c_{k_F,n} + \sum_{k_B} \hbar\omega_{k_B} c_{k_B,n}^\dagger c_{k_B,n}\right) \quad (S4)$$

$$\sum_k g_{k,m}(c_k^\dagger e^{-i\mathbf{k}\cdot\mathbf{r}_m}\sigma_m - c_k e^{i\mathbf{k}\cdot\mathbf{r}_m}\sigma_m^\dagger)$$

$$= \sum_{n=1}^N \left(\sum_{k_F} g_{k_F,m}(c_{k_F}^\dagger e^{-ik_F\xi_{m,n}}\sigma_m - c_{k_F} e^{ik_F\xi_{m,n}}\sigma_m^\dagger) \right.$$

$$\left. + \sum_{k_B} g_{k_B,m}(c_{k_B}^\dagger e^{-ik_B\xi_{m,n}}\sigma_m - c_{k_B} e^{ik_B\xi_{m,n}}\sigma_m^\dagger)\right) \quad (S5)$$

where $\omega_{k_F} = ck_F$ and $\omega_{k_B} = -ck_B$ are the dispersion relationships for forward and backward propagating photons in each channel, respectively. Note $k_F$ and $k_B$ are scalars with opposite signs.

The terms on the right-hand side of Eqs. (S4) and (S5) can be easily represented in real-space. For that purpose, we define following Fourier transformations

$$c_{k_F,n}^\dagger = \int_{-\infty}^\infty d\xi\, c_{F,n}^\dagger(\xi) e^{ik_F\xi} \quad (S6)$$

$$c_{k_B,n}^\dagger = \int_{-\infty}^\infty d\xi\, c_{B,n}^\dagger(\xi) e^{ik_B\xi} \quad (S7)$$

where $c_{F,n}^\dagger(\xi)$ and $c_{B,n}^\dagger(\xi)$ create a forward and a backward propagating photon at position $\xi$ in the $n^{th}$ channel, respectively (Fig. S1c). The summation over $k_F$ then becomes

$$\sum_{k_F} \hbar\omega_{k_F} c_{k_F,n}^\dagger c_{k_F,n} = \sum_{k_F} \hbar ck_F \int_{-\infty}^\infty \int_{-\infty}^\infty d\xi d\xi'\, c_{F,n}^\dagger(\xi) c_{F,n}(\xi') e^{ik_F(\xi-\xi')}$$

$$= \hbar c \int_{-\infty}^{\infty}\int_{-\infty}^{\infty} d\xi d\xi'\, c_{F,n}^{\dagger}(\xi) c_{F,n}(\xi') \int_{-\infty}^{\infty} \frac{k_F}{2\pi} e^{ik_F(\xi-\xi')}\, dk_F$$

$$= i\hbar c \int_{-\infty}^{\infty}\int_{-\infty}^{\infty} d\xi d\xi'\, c_{F,n}^{\dagger}(\xi) c_{F,n}(\xi') \left(-\frac{d}{d\xi}\right)\delta(\xi - \xi') \tag{S8}$$

$$= i\hbar c \int_{-\infty}^{\infty} d\xi\, c_{F,n}^{\dagger}(\xi) \left(-\frac{d}{d\xi}\right) c_{F,n}(\xi)$$

and

$$\sum_{k_F} g_{k_F,m}\left(c_{k_F}^{\dagger} e^{-ik_F \xi_{m,n}} \sigma_m - c_{k_F} e^{ik_F \xi_{m,n}} \sigma_m^{\dagger}\right)$$

$$= \sum_{k_F} g_{k_F,m}\left(\int_{-\infty}^{\infty} d\xi\, c_{F,n}^{\dagger}(\xi) e^{ik_F(\xi-\xi_{m,n})}\sigma_m - \int_{-\infty}^{\infty} d\xi\, c_{F,n}(\xi) e^{-ik_F(\xi-\xi_{m,n})}\sigma_m^{\dagger}\right)$$

$$= g_{m,n}\left(\int_{-\infty}^{\infty} d\xi\, c_{F,n}^{\dagger}(\xi)\sigma_m \int_{-\infty}^{\infty} \frac{e^{ik_F(\xi-\xi_{m,n})}dk_F}{2\pi} - \int_{-\infty}^{\infty} d\xi\, c_{F,n}(\xi)\sigma_m^{\dagger} \frac{e^{-ik_F(\xi-\xi_{m,n})}dk_F}{2\pi}\right)$$

$$= g_{m,n} \int_{-\infty}^{\infty} d\xi\, \delta(\xi - \xi_{m,n})\left(c_{F,n}^{\dagger}(\xi)\sigma_m - c_{F,n}(\xi)\sigma_m^{\dagger}\right) \tag{S9}$$

Similarly, the summation over $k_B$ becomes

$$\sum_{k_B} \hbar \omega_{k_B} c_{k_B,n}^{\dagger} c_{k_B,n} = i\hbar c \int_{-\infty}^{\infty} d\xi\, c_{B,n}^{\dagger}(\xi)\left(\frac{d}{d\xi}\right) c_{B,n}(\xi) \tag{S10}$$

and

$$\sum_{k_B} g_{m,n}\left(c_{k_B}^{\dagger} e^{-ik_B \xi_{m,n}} \sigma_m - c_{k_B} e^{ik_B \xi_{m,n}} \sigma_m^{\dagger}\right)$$

$$= g_{m,n} \int_{-\infty}^{\infty} d\xi\, \delta(\xi - \xi_{m,n})\left(c_{B,n}^{\dagger}(\xi)\sigma_m - c_{B,n}(\xi)\sigma_m^{\dagger}\right) \tag{S11}$$

Substituting Eqs. (S2) – (S11) into the Hamiltonian in Eq. (S1) we can easily obtain the Hamiltonian in real space as

$$H = \sum_{m=1}^{M} \hbar \omega_m \sigma_m^{\dagger}\sigma_m + i\hbar c \sum_{n=1}^{N} \int_{-\infty}^{\infty} d\xi\, \left(c_{F,n}^{\dagger}(\xi)\left(-\frac{d}{d\xi}\right)c_{F,n} + c_{B,n}^{\dagger}(\xi)\left(\frac{d}{d\xi}\right)c_{B,n}(\xi)\right)$$

$$+ i\hbar \sum_{m=1}^{M}\sum_{n=1}^{N} \int_{-\infty}^{\infty} d\xi\, g_{m,n}\delta(\xi - \xi_{m,n})\left\{\left(c_{F,n}^{\dagger}(\xi) + c_{B,n}^{\dagger}(\xi)\right)\sigma_m - \left(c_{F,n}(\xi) + c_{B,n}(\xi)\right)\sigma_m^{\dagger}\right\} \tag{S11}$$

Next we consider the scattering process of single photons in each channel, which is same as a 1D scattering problem in waveguide QED. In the scattering process in each channel, the incident photons pass through a chain of $M$ TLSs. Thereby there are three distinct regions for the scattered photons: the reflected, the inter-TLSs and the transmitted regions. In the inter-TLSs region, the photons are scattered back and forward between TLSs, inducing the collective interaction between TLSs. Therefore, the single-photon wavefunctions of the scattered photons in the $n^{th}$ channel then can be written as

$$\psi_{F,n}(\xi) = e^{ik\xi}\left[F_n \theta(\xi_{1,n} - \xi) + \sum_{m=1}^{M-1} C_{m,n}\theta(\xi - \xi_{m,n})\theta(\xi_{m+1,n} - \xi) + t_n \theta(\xi - \xi_{M,n})\right] \quad \text{(S13)}$$

$$\psi_{B,n}(\xi) = e^{-ik\xi}\left[r_n \theta(\xi_{1,n} - \xi) + \sum_{m=2}^{M} D_{m,n}\theta(\xi - \xi_{m-1,n})\theta(\xi_{m,n} - \xi) + B_n \theta(\xi - \xi_{M,n})\right] \quad \text{(S14)}$$

and are schematically plotted in Fig. S1d. Here $\theta(\xi)$ is the step function and $\xi_{m,n}$ is the location of the $m^{th}$ TLS projected along the $n^{th}$ channel. The coefficients $C_{m,n}$ and $D_{m,n}$ are the amplitudes of forward and backward propagating photons between the $m^{th}$ and $(m+1)^{th}$ TLSs in the $n^{th}$ channel, respectively. They satisfy the boundary condition at the location of the TLSs. The coefficients $F_n$ and $B_n$ are the incident amplitudes of the incident photon in the forward and backward directions. The coefficients $r_n$ and $t_n$ are the amplitudes of the reflected and transmitted photons by the metasurface in the $n^{th}$ channel, respectively.

Given the single photon wavefunctions in each channel, the most general eigenstate of the Hamiltonian in Eq. 1 then can be written as

$$|\psi\rangle = \sum_{n=1}^{N} \int d\xi \left[\psi_{F,n}(\xi) c_{F,n}^\dagger(\xi) + \psi_{B,n}(\xi) c_{B,n}^\dagger(\xi)\right]|0,g\rangle + \sum_{m=1}^{M} e_m \sigma_m^\dagger |0,g\rangle \quad \text{(S15)}$$

where $|0,g\rangle$ indicates that all TLSs are in ground state and $e_m$ is the excitation amplitude of the $m^{th}$ TLS.

Now we consider the solution to the Hamiltonian in Eq. (S12). Here we only discuss the steady-state solution by solving the time-independent Schrodinger equation. The transient scattering process can be solved similarly by solving the time-dependent Schrodinger equation.

By using the time-independent Schrodinger equation $H|\psi\rangle = \hbar\omega|\psi\rangle$, we can get the following equations for the forward propagating photons in each channel

$$-ice^{ik\xi_{1,n}}(-F_n + C_{1,n}) + ig_{1,n}e_1 = 0 \quad \text{(16a)}$$

$$-ice^{ik\xi_{m,n}}(-C_{m-1,n} + C_{m,n}) + ig_{m,n}e_m = 0, m = 2, \ldots, M-1 \quad \text{(16b)}$$

$$-ice^{ik\xi_{M,n}}(-C_{M-1,n} + t_n) + ig_{M,n}e_M = 0 \quad \text{(16c)}$$

Similarly, we can get the following equations for the backward propagating photons

$$ice^{-ik\xi_{1,n}}(-r_n + D_{1,n}) + ig_{1,n}e_1 = 0 \tag{17a}$$

$$ice^{-ik\xi_{m,n}}(-D_{m-1,n} + D_{m,n}) + ig_{m,n}e_m = 0, m = 2, \ldots, M-1 \tag{17b}$$

$$ice^{-ik\xi_{M,n}}(-D_{M-1,n} + B_n) + ig_{M,n}e_M = 0 \tag{17c}$$

Lastly, we can get the following equation for the excitation amplitude $e_m$ of the $m^{\text{th}}$ TLS

$$\omega_m e_m - i \sum_{n=1}^{N} g_{m,n} \left( e^{ik\xi_{m,n}} \left( \frac{F_n}{2} + \sum_{l=1}^{M-1} \frac{C_{l,n}}{2} + \frac{t_n}{2} \right) + e^{-ik\xi_{m,n}} \left( \frac{r_n}{2} + \sum_{l=1}^{M-1} \frac{D_{l,n}}{2} + \frac{B_n}{2} \right) \right) \tag{S18}$$
$$= \omega e_m$$

In order to solve above equations, we first simplify Eq. S18 by combining Eqs. (S16) – (S18) and obtain

$$(\omega - \omega_m + i\Gamma_m)e_m + i \sum_{\substack{l=1 \\ l \neq m}}^{M} \Gamma_{ml} e_l = -i \sum_{n=1}^{N} g_{m,n}(F_n e^{ik\xi_{m,n}} + B_n e^{-ik\xi_{m,n}}) \tag{S19}$$

where $\Gamma_m = \sum_{n=1}^{N} g_{m,n}^2/2c$ is the spontaneous emission rate of the $m^{\text{th}}$ TLS and $\Gamma_{ml} = \sum_{n=1}^{N} g_{m,n} g_{l,n} e^{ik(\xi_{m,n} - \xi_{l,n})}/2c$ describes the collective interaction between the $m^{\text{th}}$ and $l^{\text{th}}$ TLSs. By directly solving Eq. S19 we obtain the excitation amplitude $e_m$ of each individual TLS. The amplitudes $C_{m,n}$, $D_{m,n}$, $r_n$ and $t_n$ of the scattered photons in each channel then can be computed from Eqs. (S16) – (S17), allowing us to calculate the spatial wavefunctions of the scattered photons. Exact analytical solution can be obtained for several TLSs. For atomic metasurfaces containing more than 1000 TLSs, we obtain the solution numerically.

## 2. Creating phase gradient by the interference of multiple control beams

In this part, we discuss the details of using multi-beam interference to create the phase gradient for metasurfaces.

We use the interference of 3 control laser beams to create the structured illumination that is spatially varying along the $x$-axis. The intensity of the illumination controls the energy levels of the dressed states. Figure S2a shows the directions of the control laser beams. To create a phase gradient of $\Delta k$, the wave vectors of the control laser beams in the $x$ direction satisfy $k_x = 0, -\Delta k, 2\Delta k$. The total electric field distribution at $x$-$y$ plane is given by

$$A_c(x) = \frac{A_0}{2} + \frac{A_0}{2\pi} \sin(-\Delta k x) + \frac{A_0}{4\pi} \sin(2\Delta k x) \tag{S20}$$

where the electric field distributions of Beam-0, 1, and 2 are given by the first, second and third term, respectively. $A_0$ is the magnitude of the electric field that corresponding to a Rabi frequency of $2\Gamma$. The Rabi frequency shares the same spatial shape with the electric field. As we discussed in the main text, we set $E_{ph} = E_2 - E_1 - \Gamma/2$ so that the scattering phases of an individual atom varies approximately linearly with the Rabi frequency. Therefore, the spatially varying Rabi frequency produced by the control lasers approximates a linear phase gradient of $\Delta k$ (Fig. S2b).

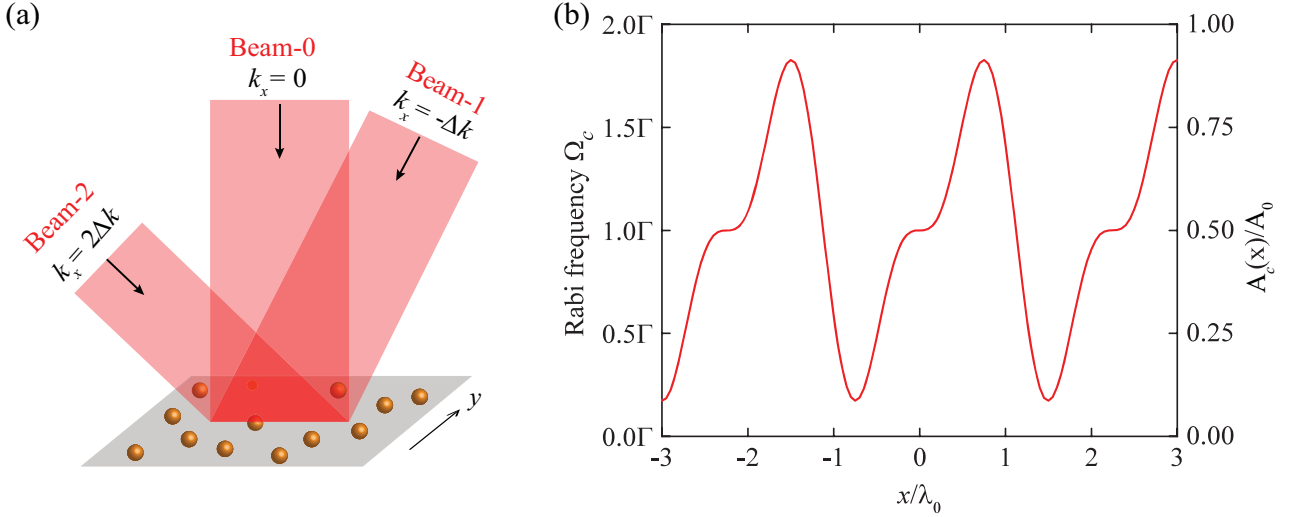

FIG. S2. (a) Schematic of the control laser beams. (b) Spatially varying electric field and Rabi frequency created by the control laser beams.

### 3. Realization of $2\pi$ phase range

A single resonance provides a scattering phase with a $\pi$ phase range. However, a $2\pi$ phase range is required to obtain maximum steering efficiency. This $2\pi$ phase range can be realized in atomic metasurfaces by directly applying one of several approaches developed for conventional metasurfaces, including cross-polarized radiation [3], Pancharatnam-Berry (PB) phase [4], as well as the use of simultaneous electric and magnetic resonances [5]. In this section, we will discuss the strategies to realize a $2\pi$ phase range by using multiple TLSs and PB phase.

Figure S3a shows a composite scattering element consisting of two TLSs that can create a $2\pi$ phase range. The two TLSs have the same transition frequency $\omega_{\text{TLS}}$ but different orientations of the transition dipole moment. The dipole moment of the first TLS (golden sphere) is along the *x*-axis and that of the second TLS (gray sphere) is along *y*-axis. They are separated by a spacing of $0.05\lambda_{ph}$, where $\lambda_{ph}$ is the wavelength of the incident single photons.

At such a small spacing, there exists a strong optical near-field interaction between the two TLSs, which is often referred as static dipole-dipole interaction [1]. The strength of the near-field interaction is given by [1]

$$\Omega_{12} = -\frac{3}{4}\Gamma(\hat{x}\cdot\hat{r}_{12})\cdot(\hat{y}\cdot\hat{r}_{12})\left(\frac{\sin k_0 r_{12}}{(k_0 r_{12})} - 3\frac{\sin k_0 r_{12}}{(k_0 r_{12})^2} + 3\frac{\cos k_0 r_{12}}{(k_0 r_{12})^3}\right) \tag{S21}$$

where $\Gamma$ is the energy bandwidth of the TLSs, $\hat{r}_{12}$ is the unit vector from TLS 1 to TLS 2, $r_{12}$ is the distance between them and $k_0$ is the wave number of the incident single photons. The strong near-field interaction creates two resonances: the subradiant and the superradiant modes. Each resonance provides a π phase range as shown in Fig. S3a, which shows the scattering phase shift (black solid line) and the scattering amplitude (red solid line) of the cross-polarized scattered photons.

Besides the phase range of 2π, the scattering amplitude also is very important to realize more advance optical functionalities. In conventional metasurfaces, the scattering elements are carefully chosen to have the same scattering amplitude [3]. Such design constrains unfavorably limit the performance of the metasurface. In great contrast, our two-TLS scattering element inherently overcome the requirement of uniform scattering amplitude. We provide such an example of two-TLS metasurface with a steering efficiency of 98.1% in Section 4. Multi-band operation based on such metasurface is also demonstrated in Section 5.

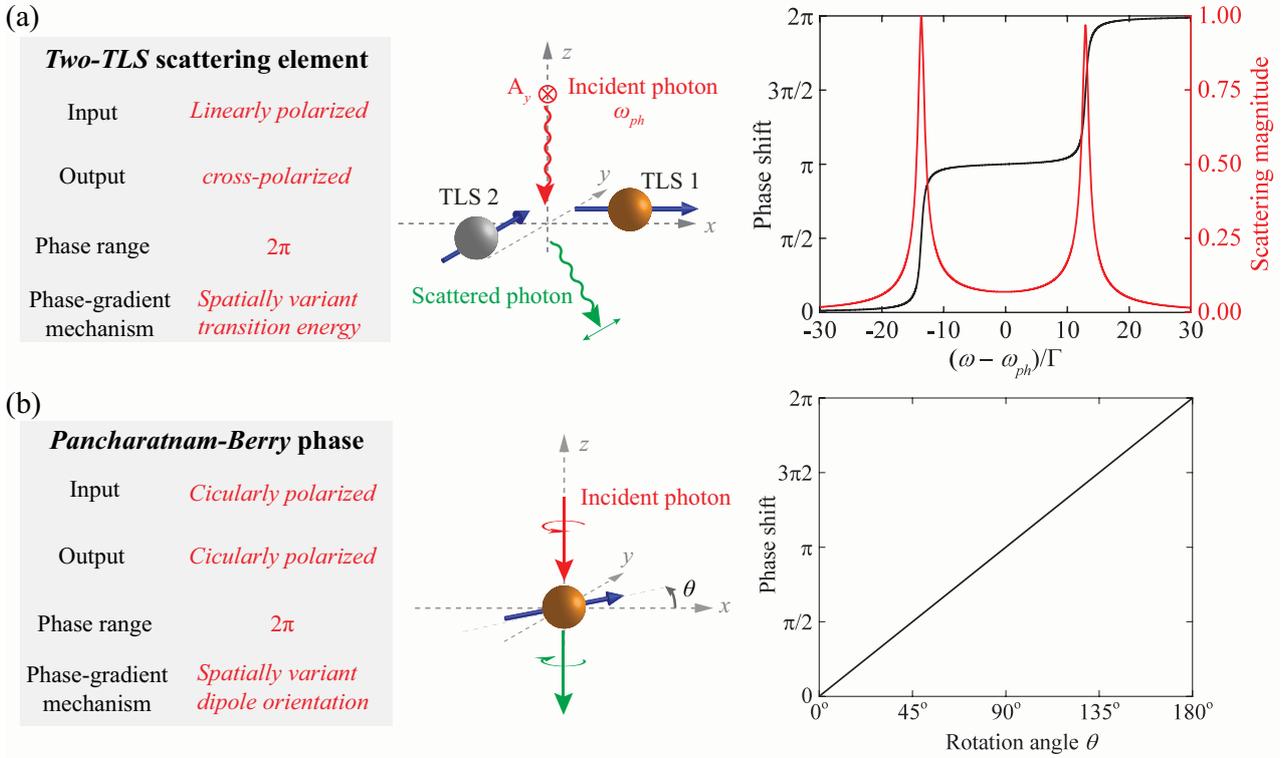

FIG. S3. (a) Two-TLS scattering element. A two-TLS scattering element (center panel) consists of two TLSs that have orthogonal dipole moments (blue arrows). The scattering phase (black solid line) and scattering magnitude (red solid line) of scattered photons are plotted in the right panel. The strong near-field interaction between the TLSs creates a subradiant resonance (narrow peak) in addition to the superradiant resonance (broad peak). As a consequence, a 2π phase range is achieved. (b) PB phase created by TLS. When circularly polarized photon interacts with a TLS, the TLS imparts a geometrical phase to the transmitted photon (center panel). This phase is determined by the orientation of the TLS in the $x$-$y$ plane. When it rotates 180 degrees, a 2π phase range is achieved (right panel).

Another way to realize a 2π phase range is using PB phase, which is created when circular polarized light interacts with a TLS (Fig. S3b). The TLS functions as a half-wave plate that flips the rotation direction of the circularly polarized light. As show in the center panel in Fig. S3b, the TLS has a permanent direction for its dipole moment (blue arrow), which can be found in solid state TLSs. Its transition energy is detuned from the incident photon's frequency (for example by 2Γ). Depending on the orientation of the TLS in the *x-y* plane, a geometrical phase from 0 to 2π can be acquired for the transmitted light (right panel in Fig. S3b), which is also circularly polarized. A metasurface can be formed by arrays of TLSs with spatially varying orientations.

## 4. Overcoming the issue of non-uniform scattering amplitude to realize high-efficiency (~ 98.1%) steering of single photons

Here we demonstrate the beam steering with 98.1% efficiency. It uses the two-TLS scattering element as described in Section. 3.

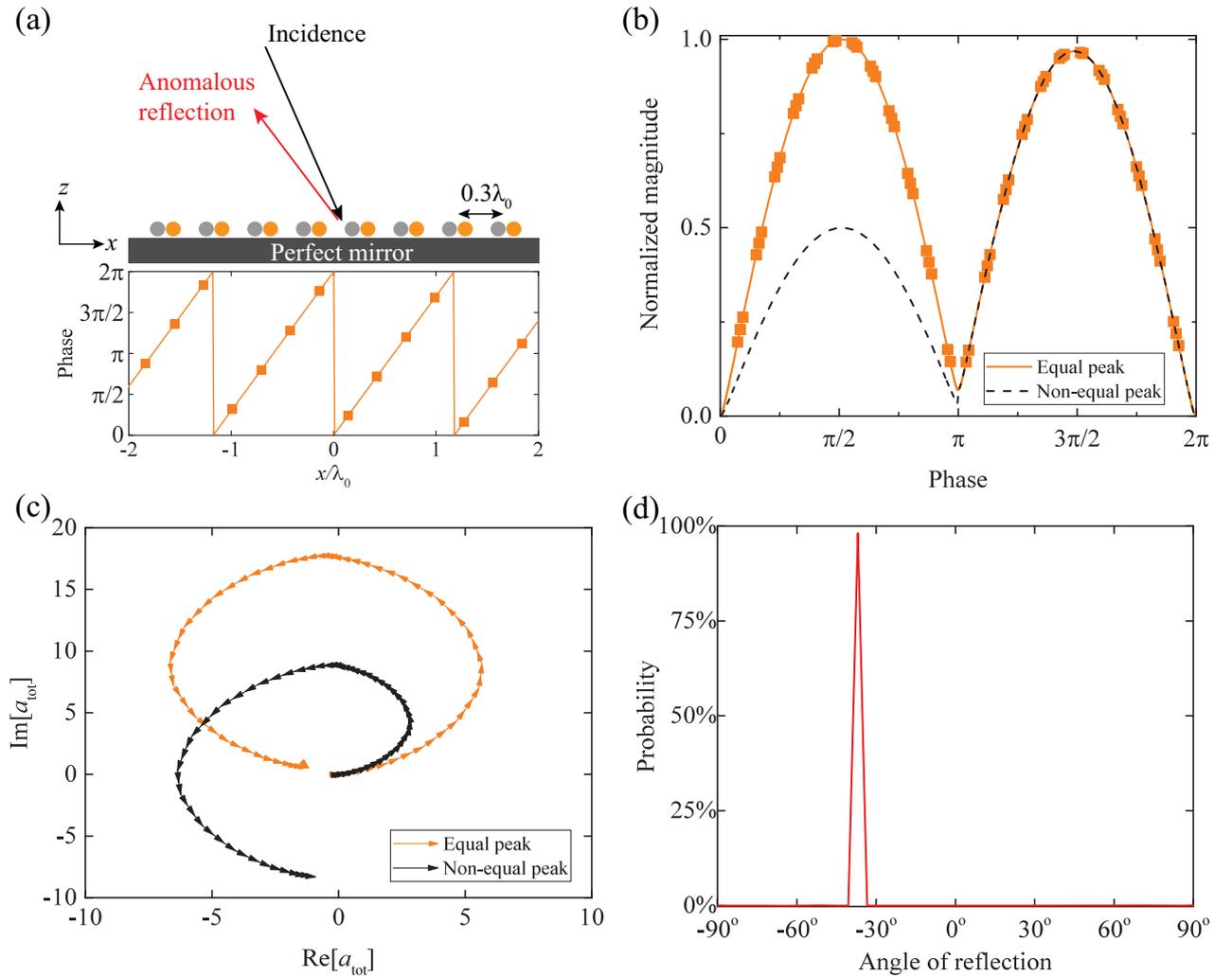

FIG. S4. (a) Schematic of an atomic metasurface based on two-TLS scattering elements. The scattering phase of each pair varies with position, which is indicated by the orange square markers in the lower panel. The scattering amplitude (black star markers) is optimized to be almost uniform everywhere. (b) Normalized scattering magnitude as a function of scattering phase. The orange solid line indicates that the superradiant and subradiant resonances have equal peak magnitude, and the black dashed line indicates that they have non-equal peak magnitude. The square markers represent the scattering elements shown in (a). (c) Phase diagram corresponding to (b). With equal peak of superradiant and subradiant resonances, the orange vectors form a closed loop, indicating complete elimination of ordinary reflection. In great contrast, non-equal peak magnitude leads to open loop (black vectors) and large ordinary reflection. (d) Reflected probabilities of cross-polarized single photons. The probability to the anomalous reflection reaches 98%. The rest 2% of single photons are co-polarized and about half of them are reflected anomalously.

We design a metasurface to steer light in the *x*-axis direction. It consists of a 2-dimension array of two-TLS elements that creates a phase gradient along the *x* direction [Fig. S4a]. The spacing between elements is $0.3\lambda_0$, where $\lambda_0$ is the working wavelength. In order to remove the ordinary and anomalous refractions, we put the metasurface on a perfect mirror. The scattering phase of each element varies linearly with position, which is indicated by the orange square markers in the lower panel of Fig. S4a.

To overcome the issue of non-uniform scattering amplitudes, we only need to have the scattering amplitudes being the same at the resonant frequencies of the superradiant and subradiant resonances. To clearly demonstrate the underlying principle, we plot the scattering magnitude as a function of the scattering phase in Fig. S4b as orange solid line, where the orange square markers represent the elements in the metasurface. For the ordinary reflection, the total amplitude of wavefunction $a_{tot}$ is given by $a_{tot} = \sum_{m=1}^{M} |a_m| e^{i\phi_m}$, where $|a_m|$ and $\phi_m$ are the scattering magnitude and phase of $m$[th] element, respectively. We further plot the scattering amplitudes of the scattering elements in the complex plane, where they are represented by orange vectors [Fig. S4c]. The equal peak magnitude of the superradiant and subradiant resonances ensures that the vectors form a closed loop, indicating the total amplitude of ordinary reflection $a_{tot}$ is zero. In great contrast, if we assume the peak magnitude is non-equal [black dashed line in Fig. S4b], the black vectors in Fig. S4c will not form a closed loop, leading to large ordinary reflection.

We now calculate the reflection probabilities to validate above assumption. Incident single photons are polarized along the *y*-axis and incident obliquely onto the metasurface. The calculated probabilities of cross-polarized single photons are plotted in Fig. S4d. Here we only plot the probabilities to reflection. 98.1% of incident single photons are steered to the anomalous reflection direction. Moreover, the ordinary reflection is completely eliminated, validating our assumption.

## 5. Multi-band steering of single photons

Operation at multiple bands has been quite challenging for conventional optical elements. Although multi-band metasurfaces based on complex multi-resonant unit cells have been demonstrated recently [6,7], their design procedure is complex, and the number of bands is inversely proportional to the metasurface efficiency.

The extremely small physical size of atomic TLSs leads large empty spaces between neighboring scattering elements in the metasurface. Here we show that a multi-band metasurface can be realized by filling this "empty" space with multiple overlapping metasurfaces that are designed to work at different wavelengths. Photons with different wavelengths can be efficiently steered to the same direction.

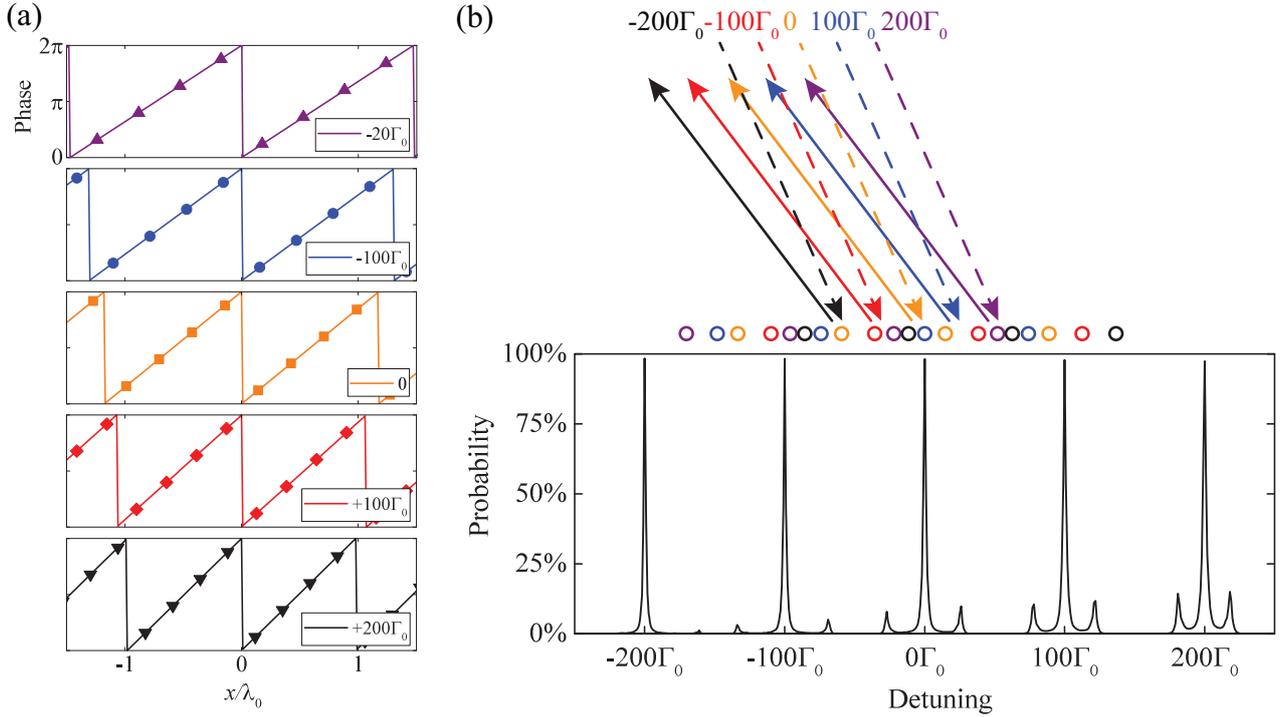

FIG. S5. (a) Scattering phase of all metasurface elements. We prepare 5 independent metasurfaces based on two-TLS scattering elements, with each metasurfaces operating at different wavelengths. The markers indicate the locations of every two-TLS element. (b) Schematic of the resulting multi-band metasurface and its spectral anomalous reflection. The metasurfaces are packed together to form a single metasurface. When single photons with different wavelength are incident from the same direction, they will be steered to the same direction. The spectral probability in the lower panel confirms the intuitive design.

Following the metasurface we demonstrated in Section. 4, we design another four independent metasurfaces for frequencies that are detuned from $\omega_0$ by $-200\Gamma_0$, $-100\Gamma_0$, $+100\Gamma_0$ and $+200\Gamma_0$, where $\Gamma_0$ of the energy bandwidth of the TLS that has a transition energy of $\omega_0$. The scattering phases at different spatial locations for each metasurface are plotted in Fig. S5a. All metasurfaces are optimized for maximal efficiency. They are also designed to have the same steering effect.

We then integrate 5 metasurfaces on a single plane to form a multi-band metasurface. With incident photons in different frequencies, the multi-band metasurface steers them to the same direction. Different frequencies are indicated by different colors in Fig. S5b. The probability of finding photons in the designed direction is calculated from quantum scattering theory. Over 98% efficiency is achieved for all 5 bands.

## 6. Phase gradient for steering of spontaneous emission of single photons

As we discussed in the main text, we specifically design a phase profile to steer the spontaneous emission of a single excited TLS within a metasurface. This phase profile is significantly different from the linear phase profile of a passive metasurface. The propagation phase of the emitted photons has to be considered. For simple design process, we will only take into account the first-order propagation phase of emitted photons. However, the beaming angle approximates the designed angle very well.

The designed spatial phase profile is plotted in Fig. S6a as the red curve. The designed beaming angle is $\theta = -77.2°$. The scattering magnitude is also encoded by the weight of the red curve. The black dashed line represents the ideal spatial phase profile with first-order correction for beaming photons to the designed angle. We specifically minimize the scattering magnitude in the region where the scattering phase cannot be covered by single TLSs. Note we also ignore the near-field interaction between TLSs since it only shifts the scattering phase of a given TLS. The corresponding transition energies of all TLSs are then determined by the phase relation described in Fig. 1c.

Figure S6b shows the total flux of probability integrated over time as red solid line. Most of the emitted photons are beamed to $\theta = -80°$, which is very close to the designed direction (black dashed arrow).

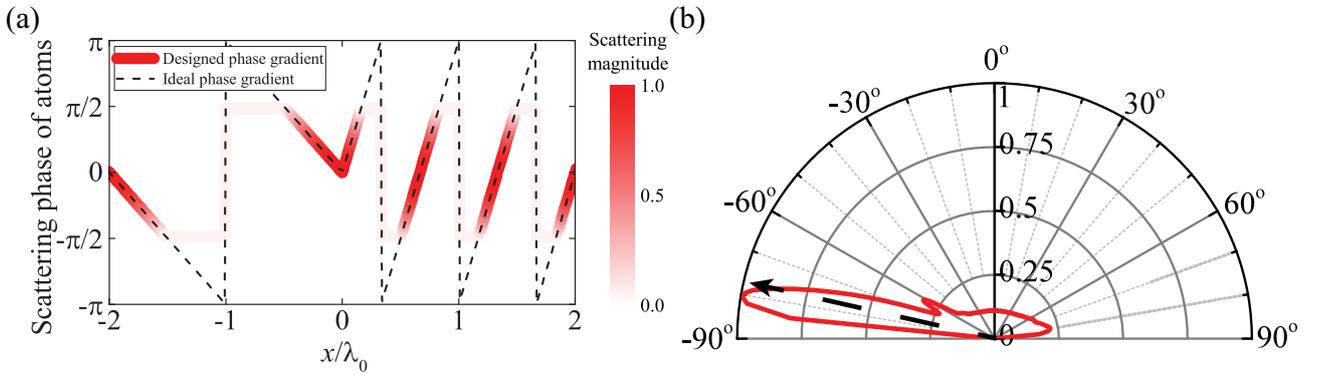

FIG. S6. (a) The red curve represents the designed spatial phase profile. The scattering magnitude is encoded by the weight of the red curve. The black dashed line represents the ideal phase profile with first-order corrections for beaming photons to the designed direction. (b) Total flux of probability integrated over time. The black dashed arrow indicates the designed direction.

**Reference**

[1] Z. Ficek and S. Swain, *Quantum Interference and Coherence: Theory and Experiments* (Springer Science & Business Media, 2005).
[2] J.-T. Shen and S. Fan, Phys. Rev. A **79**, 23837 (2009).
[3] N. Yu, P. Genevet, M. A. Kats, F. Aieta, J.-P. Tetienne, F. Capasso, and Z. Gaburro, Science **334**, 333 (2011).
[4] D. Lin, P. Fan, E. Hasman, and M. L. Brongersma, Science **345**, 298 (2014).
[5] S. Sun, K.-Y. Yang, C.-M. Wang, T.-K. Juan, W. T. Chen, C. Y. Liao, Q. He, S. Xiao, W.-T. Kung, G.-Y. Guo, L. Zhou, and D. P. Tsai, Nano Lett. **12**, 6223 (2012).
[6] M. Khorasaninejad, F. Aieta, P. Kanhaiya, M. A. Kats, P. Genevet, D. Rousso, and F. Capasso, Nano Lett. **15**, 5358 (2015).
[7] F. Aieta, M. A. Kats, P. Genevet, and F. Capasso, Science **347**, 1342 (2015).